\documentstyle[11pt]{article}
\def\bkR{{\rm I\kern-.17em R}}
\def\bkC{{\rm \kern.24em \vrule width.05em height1.4ex depth-.05ex \kern-.26em C}}
\def\bkN{{\rm \kern.50em \vrule width.05em height1.4ex depth-.05ex \kern-.26em N}}

\begin{document}

\author{Nuno Costa Dias\footnote{{Email: ncdias@mail.telepac.pt}} \\ Jo\~{a}o Nuno Prata\footnote{{Email: joao.prata@mail.telepac.pt}} \\ {\it Departamento de Matem\'atica} \\
{\it Universidade Lus\'ofona de Humanidades e Tecnologias} \\ {\it Av. Campo Grande, 376, 1749-024 Lisboa, Portugal}\\
{\it and}\\
{\it Grupo de F\'{\i}sica Matem\'atica}\\
{\it Universidade de Lisboa}\\
{\it Av. Prof. Gama Pinto 2}\\
{\it 1649-003 Lisboa, Potugal}}

\title{DEFORMATION QUANTIZATION OF CONFINED SYSTEMS\footnote{Presented by N.C. Dias at the {\it Workshop on Advances in Foundations of Quantum Mechanics and Quantum Information with atoms and photons}, 2-5 May 2006, Turin, Italy.}}

\maketitle

\begin{abstract}
The Weyl-Wigner formulation of quantum confined systems poses several interesting problems. The energy stargenvalue equation, as well as the dynamical equation does not display the expected solutions. In this paper we review some previous results in the subject and add some new contributions. We reformulate the confined energy eigenvalue equation by adding to the Hamiltonian a new (distributional) boundary potential. The new Hamiltonian is proved to be globally defined and self-adjoint. Moreover, it yields the correct Weyl-Wigner formulation of the confined system.
\end{abstract}

{\bf Keywords}: {Quantum systems with boundaries; deformation quantization; operator methods.}

\section{Introduction}
Deformation quantization is a powerful procedure to derive the quantum mechanical formulation of a wide range of physical systems. These include systems defined in non-flat Poisson or sympletic manifolds \cite{Kontsevich}. In the simplest case of flat phase spaces the deformation approach yields the Weyl-Wigner formulation of quantum mechanics \cite{Weyl}-\cite{Dias2}. This is an autonomous formulation of quantum mechanics that has been an active research topic in both physics and mathematics.

Quite unexpectedly the deformation approach displays several problems in the case of confined systems \cite{Dias3}-\cite{Dias4}. The dynamical and the eigenvalue equations do not yield the correct solutions. Given the status of the deformation method this is a serious drawback. Moreover the importance of confined systems in several domains of mathematical physics \cite{Piotr}-\cite{Bonneau} (e.g. the theory self-adjoint extensions of symmetric operators), condensed matter physics (e.g. the quantum description of particles moving on surfaces with obstacles or impurities) and even in string theory and other modern approaches to quantum gravity \cite{Isham} (where the classical theory displays a non trivial topological structure) justifies a more thorough investigation of the quantum phase space formulation of these systems.
 
In this paper we study the simplest version of this problem: the Weyl-Wigner formulation of the energy eigenvalue equation for a 1-dimensional particle confined to the negative semi-axis and subject to Dirichlet boundary conditions. 
We will identify the source of the problem at the level of the standard operator formulation of quantum mechanics by showing that the confined energy eigenfunctions are not global (i.e. distributional) solutions of the textbook eigenvalue equation. We will propose a new, globally valid eigenvalue equation which is written in terms of a new Hamiltonian containing a confining distributional potential. The translation of this equation to the Weyl-Wigner formulation yields a new stargenvalue equation displaying the correct solutions. This problem has been studied before \cite{Dias3}-\cite{Dias4}. Here we reformulate it on simpler grounds and add some new contributions. In particular we show that the new Hamiltonian is self-adjoint. From the point of view of the theory of self-adjoint extensions of symmetric operators this is also a non-trivial result \cite{Piotr}-\cite{Bonneau}.

This paper focuses on the stargenvalue equation but a similar approach is valid for the dynamical equation. The method presented here is extendable to generic Hamiltonians of the form $H=\frac{p^2}{2m} +V(q)$ and to general boundary conditions. 
 
\section{The Weyl-Wigner formulation of confined systems}

Let us consider the Gelf'and triple ${\cal G} \subset {\cal H}=L_2(\bkR) \subset {\cal G'}$ where ${\cal G}$ is the space of infinitely smooth functions $t(x)$ that, together with all its derivatives, decay to zero faster than any power of $1/x$ as $|x| \to \infty$, ${\cal G}'$ is its dual and $L_2(\bkR)$ is the space of complex valued square integrable functions on $\bkR$. Generalized operators are linear maps from ${\cal G} $ to ${\cal G'}$. 
The set of these operators is denoted by $\hat{\cal T}$ and (in a suitable sense) it contains the sets of Hilbert-Schmidt operators over ${\cal H}$ as well as the enveloping algebra of the Heisenberg-Weyl Lie algebra. The Weyl-Wigner map is a one to one invertible transformation from $\hat{\cal T}$ onto the set of generalized functions on $T^*M \simeq \bkR^2$ (we make $\hbar =1$) \cite{Bracken}:  
\begin{equation}
 \hat A \longrightarrow A(x,p) \equiv W[\hat A]= \int dy \, e^{-ipy} <x+y/2|\hat A |x-y/2>.
\end{equation}
This map provides the translation of all the important structures of operator quantum mechanics into the quantum phase space language: operators are mapped to phase space functions (denoted by symbols) $A(x,p)=W(\hat A)$, and the density matrix to the celebrated Wigner quasi-distribution
\begin{equation}
f_W(x,p) = \frac{1}{2\pi} W(|\psi><\psi|)=\frac{1}{2\pi} \int dy \, e^{-ipy} <x+y/2|\psi><\psi |x-y/2> .
\end{equation}
The operator product and bracket yield the algebraic structures of the non-commutative phase space. We have
\begin{equation}
A(x,p) \star B(x,p) = \left. \exp \left[\frac{i}{2} \left( \frac{\partial}{\partial x}\frac{\partial}{\partial p'}-\frac{\partial}{\partial p}\frac{\partial}{\partial x'} \right) \right] A(x,p) B(x',p') \right|_{(x',p') = (x,p)},  
\end{equation} 
for the $\star$-product and $\left[A , B\right]_M= -i(A\star B -B \star A )$ for the Moyal bracket.
The two key equations of the Weyl-Wigner formulation are the Moyal dynamical equation and the $\star$-genvalue equation. The former reads
\begin{equation}
A \star F_{aa} = F_{aa} \star A =a F_{aa}.
\end{equation}
It amounts to the Weyl transform of the eigenvalue equation of the standard operator formulation and is of equal importance in the context of quantum phase space methods. Its solutions (the $\star$-genfunctions) are $\hbar$-deformations of the Dirac measure, being central to the calculation of probability distributions. 

Let us now consider a one dimensional quantum particle described by the Hamiltonian ($\hbar = 2m=1$)
\begin{equation}
\hat H_0=\hat p^2= -\frac{\partial^2}{\partial x^2},
\end{equation}
which is confined to the negative semi-axis $x\le 0$ and subject to Dirichlet boundary conditions at $x=0$.
The energy eigenstates of the system are\footnote{$\phi_E(x)$ are the free eigenfunctions and $\theta(x)$ is the Heaviside step function.}
\begin{equation}
\psi_E(x)=\theta (-x) \phi_E(x)=\theta(-x) \left[e^{i\sqrt{E} x} -e^{-i\sqrt{E} x} \right], \qquad E>0.
\end{equation}
The Weyl-Wigner transform of which are
\begin{equation}
\begin{array}{c}
F_{EE}(x,p)= \theta(-x) \left[\frac{2\sin \left[2x(p+\sqrt{E}) \right]}{p+\sqrt{E}}+
 \frac{2\sin \left[2x(p-\sqrt{E}) \right]}{p-\sqrt{E}}
+\frac{4}{p}\cos (2x\sqrt{E})\sin(2xp) \right].
\end{array}
\end{equation}
Contrary to our expectations these phase space functions do not satisfy the energy stargenvalue equations (not even for $x\le 0$)
\begin{equation}  
H \star F_{EE} \not=  E F_{EE} \quad \mbox{and} \quad 
F_{EE} \star H \not= EF_{EE}.
\end{equation}
The questions we want to answer are the following: (i) Why aren't the stargenfunctions solutions of the stargenvalue equation?
(ii) Is it possible to modify the Hamiltonian so that it generates the proper solutions? (iii) Is the new Hamiltonian self-adjoint?

To look for some guidance let us go back to the operator formulation of quantum mechanics. Substituting the eigenfunctions (6) in the eigenvalue equation we easily find that they are not global eigenstates of (5).
In fact
\begin{equation}
\hat H_0 \psi_E(x) =  -\frac{\partial^2}{\partial x^2} \left[\theta (-x) \phi_E(x)\right] =  E \psi_E(x) +2 \delta(x)  \phi_E'(x) + \delta'(x)  \phi_E(x).
\end{equation}
The new distributional term on the rhs is not zero for Dirichlet boundary conditions. 
Hence, the energy eigenvalue equation is also not globally valid for confined systems.
Since the Weyl-Wigner map applies to the set of generalized operators it comes as no surprise that eqs.(8) are not valid either.

\section{Global formulation of the quantum confined particle}

We now propose an alternative formulation of the confined energy eigenvalue equation which: (i) Is valid in distributional sense, i.e. the confined eigenfunctions are generalized solutions of the new eigenvalue equation, (ii) is formulated in terms of a Hamiltonian displaying a new confining boundary potential, and such that (iii) the new Hamiltonian is self-adjoint.

The first step is the introduction of the following definitions:\\
{\bf Definition 1: Special distributions of order $n$}

{\it A Schwartz distribution\footnote{The space ${\cal D}'$ of Schwartz distributions \cite{Zemanian,Hormander} is the dual of the space ${\cal D}$ of infinitely smooth functions of compact support.} $F$ is a special distribution of order $n$ iff for each $x_0 \in \bkR$ there is a left and a right neighborhood of $x_0$, $\Omega_-=]x_0-\sigma_-,x_0[$ and $\Omega_+=]x_0,x_0+\sigma_+[$, ($\sigma_-,\sigma_+>0$), and two $C^{n}(\bkR)$-functions $f_-$ and $f_+$ such that, in the sense of distributions, "$F=f_-$ on $\Omega_-$" and "$F=f_+$ on $\Omega_+$". The set of points where $F$ is not a $C^n(\bkR)$-function is denoted by} sp supp$^n F$ {\it and the set of special distributions of order $n$ by ${\cal S}^n$. In particular we have} sp supp$^{\infty}F=$sing supp $F$ \cite{Zemanian}. 

This definition provides a hierarchy of spaces ${\cal S}^{\infty} \subset 
...\subset  {\cal S}^1 \subset  {\cal S}^0 $.  
Trivial examples of elements of ${\cal S}^{\infty}$ are the Dirac delta $\delta$ and all its derivatives and of ${\cal S}^0$ are all limited functions which are continuous except on a finite subset of $\bkR$.\\
{\bf Definition 2: The regularized delta operators $\hat{\delta}^{(n)}_{\pm}(x-a)$, $a \in \bkR$}
 
{\it $\hat{\delta}^{(n)}_{\pm}(x-a)$ are the linear operators
\begin{eqnarray}
\hat{\delta}^{(n)}_{\pm}(x-a) & : & {\cal S}^n\longrightarrow {\cal S}^{\infty}; \nonumber \\
&& F \longrightarrow \hat{\delta}^{(n)}_{\pm}(x-a) \left[F\right] =\lim_{\epsilon \to 0^+} {\delta}^{(n)}(x-a) \cdot F(x\pm \epsilon).
\end{eqnarray}  
where the product $\cdot $ appearing on the rhs is the one proposed by L. H\"ormander in Ref.\cite{Hormander}.}

Notice that for $\epsilon \in ]0,\sigma[$ and $\sigma $ sufficiently small we have sing supp ${\delta}^{(n)}(x-a) \cap$ sp supp$^nF(x \pm \epsilon) = \{a\} \cap $ sp supp$^nF(x \pm \epsilon) =\emptyset$. Hence the product of distributions on the rhs of eq.(10) is well defined in the sense of the product of distributions with non-intersecting special supports proposed in \cite{Hormander}. This product yields $\delta^{(n)}(x-a) f_{\pm}(x\pm \epsilon)$, where $f_{\pm}$ is the $C^{n}$-function associated to $F$ in the right (left) neighborhood of $a$ (cf. Definition 1). Consequently, the limit of (10) exists in ${\cal S}^{\infty}$.

Let us illustrate the action of these operators with two simple examples
\begin{equation}
\hat{\delta}^{(n)}_{+}(x) \left[\theta ^{(m)}(x)\right] = \delta_{m,0} {\delta}^{(n)}(x) \qquad \mbox{and} \qquad
\hat{\delta}^{(n)}_{-}(x) \left[\theta ^{(m)}(x)\right] =0.
\end{equation}

Using the operators $\hat{\delta}^{(n)}_{\pm}$ we can re-write the energy eigenvalue equation (9) as
\begin{equation}
-\frac{\partial^2}{\partial x^2} \psi_E(x)= -\frac{\partial^2}{\partial x^2} \left[\theta (-x) \phi_E(x) \right] 
= E \psi_E(x) +2 \hat{\delta}_-(x)  \psi_E'(x) + \hat{\delta}_-'(x)  \psi_E(x).
\end{equation}
Notice that for $\psi(x<0)=\phi(x<0)$ and $\phi(x) \in C^{\infty}(\bkR) $ we have, in the distributional sense, $2 \hat{\delta}_-(x)  \psi'(x) + \hat{\delta}_-'(x)  \psi(x)=2 \delta(x)  \phi'(x) + \delta'(x)  \phi(x)$.
Imposing the Dirichlet boundary conditions $\psi_E(0)=0$, eq.(12) reduces to
\begin{equation}
-\frac{\partial^2}{\partial x^2} \psi_E(x)
= E \psi_E(x)  -\hat{\delta}_-'(x)  \psi_E(x).
\end{equation}
Hence, the new (confining) Hamiltonian is
\begin{equation} 
\hat H=\frac{\hat p^2}{2m} + \frac{\hat{\delta}_-'(x)}{2m}=\hat H_0+\frac{\hat{\delta}_-'(x)}{2m}.
\end{equation}
We can solve the new eigenvalue equation (13), both analytically and numerically (with a suitable smooth regularization of the Dirac delta) \cite{Dias3}, for Dirichlet boundary conditions and show that all its solutions are of the form (6).    

To put this formulation on firm mathematical grounds, we now specify the domain of the Hamiltonian and prove that it is self-adjoint. 
Let us assume the Hilbert space to be $
{\cal H}_c \equiv \{\psi \in L_2(\bkR) : \psi(x) = \theta(-x) \phi(x) \} $
and consider the action of $\hat H$. Its maximal domain (i.e. the largest domain where $\hat H$ is well defined) is\footnote{$AC^2(\bkR)$ denotes the set of functions defined in $\bkR$ with absolutely continuous first derivative.}
\begin{equation}
{\cal D}_{max} = \{\psi \in {\cal H}_c : \psi = \theta(-x) \phi(x), \, \phi(x) \in AC^2(\bkR) ,\, \phi(0)=0 \}.
\end{equation}
To prove this let us consider the action of $\hat H $ on a generic $\psi(x) = \theta(-x) \phi(x) $. We have
\begin{eqnarray}
\hat H \psi(x) &=& -\left[ \theta(-x) \phi''(x)-2\delta(x) \phi'(x) -\delta'(x)\phi(x) \right] + \hat{\delta}'_-(x)\psi(x) \nonumber \\
&=& -\theta(-x) \phi''(x)+ \left[ 2\delta(x) \phi'(x) +2\delta'(x)\phi(x) \right],
\end{eqnarray}
and so 
\begin{equation}
\hat H \psi(x) \in {\cal H}_c \Longrightarrow   \delta(x) \phi'(x) +\delta'(x)\phi(x)=0 \Longrightarrow \phi(0) =0.
\end{equation}
Hence $\hat H$ acts as
\begin{equation}
\hat H:{\cal D}_{max} \longrightarrow {\cal H}_c ;\quad  \hat H \left[ \theta(-x)\phi(x)\right] =-\theta(-x) \phi''(x) .
\end{equation}
Notice that ${\cal D}_{max}$ is {\it dense} in ${\cal H}_c$ and that $\hat H$ is {\it closed} in ${\cal D}_{max}$. 
To study the self-adjointness of $\hat H$ let us introduce the sesquilinear form
\begin{equation}
w_*(\xi,\psi)= (\xi,\hat H \psi)- (\hat H \xi,\psi)= \int_{-\infty}^{+\infty} dx  \overline{\xi} \hat H \psi -\psi \overline{\hat H \xi}, \quad 
\xi, \psi \in {\cal D}_{max},
\end{equation} 
and we easily find that
\begin{equation}
w_*(\xi,\psi)=\overline{\xi'(0)}\psi(0) - \overline{\xi(0)} \psi'(0) = 0 ,\quad \forall \xi,\psi \in {\cal D}_{max},
\end{equation}
and so $\hat H$ is symmetric in its maximal domain. Therefore $\hat H$ is also self-adjoint.

To check this result we may use von Neumann's theorem on deficiency indices. For $x <0$ the solutions of
\begin{equation}
(\hat H \pm ik^2) \psi =0,\quad k>0,
\end{equation}
which belong to ${\cal H}_c$, are $
\psi_{\pm} (x)= C_{\pm} \exp \left[\frac{k\pm ik}{\sqrt{2}}x \right] $
and none (except for the trivial solution) satisfies the Dirichlet boundary conditions at $x=0$. Hence, the deficiency indices of $\hat H$ are $(0,0)$ meaning that $\hat H$ is indeed self-adjoint.

\section{Wigner formulation of confined systems}

Using the Hamiltonian (14) we may now write the new stargenvalue equation
\begin{equation}
\left[\frac{p^2}{2m} + \frac{1}{2m} \delta'_- (x)\right] \star F_{EE}=E F_{EE}  \quad \mbox{and} \quad F_{EE} \star \left[\frac{p^2}{2m} + \frac{1}{2m} \delta'_- (x)\right]=E F_{EE}, 
\end{equation}
where the distributional term is evaluated according to
\begin{equation} 
\delta'_-(x) \star F_{EE}(x,p) = \lim_{\epsilon \to 0^+} \left[\delta'(x) \star F_{EE}(x-\epsilon,p) \right],
\end{equation}
and impose the boundary conditions:
\begin{equation}
\psi_E(0)=0 \Longleftrightarrow \overline{\psi_E(0)} \psi_E(0) =0 \Longleftrightarrow \int dp \,F_{EE}(0,p)=0. 
\end{equation}  
Equation (22) with the boundary conditions (24) displays the unique solutions (7). This was proven in \cite{Dias3}.

\section*{Acknowledgments} 

This work was partially supported by the grants POCTI/MAT/45306/2002 and POCTI/0208/2003 of the Portuguese Science Foundation.

\end{document}